\documentclass[12pt]{article}
\usepackage{amsmath,amsfonts,amssymb}

\def\tPi{\tilde{\Pi}}

\def\ua{\underline{a}}
\def\ub{\underline{b}}

\def\bB{\mathbf{B}}

\def\be{\begin{equation}}

\def\ee{\end{equation}}

\def\bea{\begin{eqnarray}}

\def\eea{\end{eqnarray}}

\def\bh{\bar{h}}

\def\mH{\mathcal{H}}

\newcommand{\mG}{\mathcal{G}}

\newcommand{\bT}{\mathbf{T}}

\newcommand{\mL}{\mathcal{L}}

\def\pb #1{\left\{#1\right\}}

\begin{document}

	\begin{titlepage}

		\vskip 0.4 cm
		
		\begin{center}
			{\Large{ \bf Canonical Analysis of New Non-Relativistic
					String Action and  Uniform Light-Cone Gauge Formulation
			}}
			
			\vspace{1em}  
			
			\vspace{1em} J. Kluso\v{n} 			
			\footnote{Email addresses:
			 klu@physics.muni.cz (J.
				Kluso\v{n}) }\\
			\vspace{1em}
%			\textit{Department of Physics, University of Helsinki,
%				P.O. Box 64,\\ FI-00014 Helsinki, Finland}\\
%			\vspace{.3em} $^b$\textit{Department of Physics, Tafresh University, Tafresh, Iran}\\
%			\vspace{.3em} 
\textit{Department of Theoretical Physics and
				Astrophysics, Faculty of Science,\\
				Masaryk University, Kotl\'a\v{r}sk\'a 2, 611 37, Brno, Czech Republic}
			
			\vskip 0.8cm
			
%			
%			
%			Josef Kluso\v{n}$\,^1$
%			\footnote{Email address:
%				klu@physics.muni.cz}\\
%			\vspace{1em} $^1$\textit{Department of Theoretical Physics and
%				Astrophysics, Faculty
%				of Science,\\
%				Masaryk University, Kotl\'a\v{r}sk\'a 2, 611 37, Brno, Czech Republic}\\
%			
%			
			%{\large Josef Kluso\v{n}$^{}$\footnote{E-mail: {\tt
			%klu@physics.muni.cz}} }
			%
			%\vskip 0.8cm
			%
			%{\it Department of
			%Theoretical Physics and Astrophysics\\
			%Faculty of Science, Masaryk University\\
			%Kotl\'{a}\v{r}sk\'{a} 2, 611 37, Brno\\
			%Czech Republic\\
			%[10mm]}
			
			\vskip 0.8cm
			
		\end{center}

		\begin{abstract}
We perform canonical analysis of new non-relativistic string action that
was found recently in 		[arXiv:2107.00642 [hep-th]]. We also discuss
its gauge fixed form.	

		\end{abstract}
		
		\bigskip
		
	\end{titlepage}
	
	\newpage

%%%%%%%%%%%%%%%%%%%%%
%%%%Introduction %%%%%%%%%
%%%%%%%%%%%%%%%%%%%%
\section{Introduction and Summary}
There are several examples of consistent two dimensional field theories
defined on  string world-sheet. One such an example is non-relativistic string theory that was originally introduced twenty years
ago in \cite{Gomis:2000bd,Danielsson:2000gi}. Renewed interest in this theory 
began with remarkable paper \cite{Andringa:2012uz} where new stringy Newton-Cartan
geometry was introduced by specific limiting procedure with foliation of space-time 
into two dimensional longitudinal subspace and transverse one where longitudinal subspace has origin in the idea that the non-relativistic space-time is probed by two dimensional object-string. This proposal was subsequently studied in many papers, see for example 
\cite{Bidussi:2021ujm,Bergshoeff:2021tfn,Hartong:2021ekg,Blair:2021ycc,Bergshoeff:2021bmc,Gomis:2020izd,
	Kluson:2020rij,Yan:2019xsf,Bergshoeff:2019pij,Harmark:2019upf,Gallegos:2019icg,
	Gomis:2019zyu,Kluson:2019uza,Kluson:2018vfd,Bergshoeff:2018vfn,Kluson:2018grx,Bergshoeff:2018yvt}. Another, seemeangly different approach for definition of non-relativistic string theory, was presented in remarkable paper 
\cite{Harmark:2017rpg} where non-relativistic string in torsional Newton-Cartan background was defined by null dimensional reduction from higher dimensional
space-time with light-like isometry, for further works see for example
\cite{Kluson:2021djs,Kluson:2021pux,Fontanella:2021hcb,Kluson:2021sym,Harmark:2020vll,Kluson:2020aoq,Hansen:2020pqs,Kluson:2019xuo,Kluson:2019avy,Hansen:2019pkl,Kluson:2018egd}. Finally the relation between these two seamingly different non-relativistic string theories was found in \cite{Harmark:2019upf}.

Recently new form of non-relativistic string theory was formulated in
\cite{Bidussi:2021ujm}. This new Newton-Cartan geometry is called as 
 torsional string Newton-Cartan (TSNC) geometry. In this formulation we can 
 see close analogy between non-relativistic limit of point particle and its coupling
 to Newton-Cartan geometry. Explicitly, in the point particle case the particle naturally couples to mass form $m_\mu$ while in case of non-relativistic string
 in (TSNC) geometry the corresponding object is two form gauge field $m_{\mu\nu}$. This idea was further elaborated in \cite{Bidussi:2021ujm} where new non-relativistic string algebra, called as F-string Galilei algebra, was found.
  The  non-relativistic string action in TSNC geometry was studied very carefully in 
 \cite{Bidussi:2021ujm} and brings new opportunities for the analysis of non-relativistic theories.

 The goal of this paper is to find canonical formulation of this theory and further study its gauge fixed version that could be potentially useful for non-relativistic
 expansion of various string theory backgrounds, as for example $AdS_5\times S^5$. As in case of ordinary string we find that the bare Hamiltonian vanishes which is a consequence of the fact that non-relativistic string is diffeomorphism invariant theory on two dimensional world-sheet. As a result we have to find  two primary constraints. It is rather easy to determine spatial diffeomorphism constraint starting from the definition of canonical momenta. The situation is more difficult in case of Hamiltonian constraint however it is still simpler than in case of non-relativistic string in stringy Newton-Cartan background 
 \cite{Kluson:2018grx}. This is again very remarkable property of the new formulation of non-relativistic string. We also show that these two constraints are first class
 constraints that prove consistency of this theory.
 
 As the next step we use this Hamiltonian formalism to formulate non-relativistic theory in the uniform light-cone gauge. Such a specific gauge  was introduced in the context of study of string theory on $AdS_5\times S^5$ background
 \cite{Frolov:2019nrr,Arutyunov:2006gs,Frolov:2006cc,Arutyunov:2005hd}
  \footnote{For review, see for example \cite{Arutyunov:2009ga}.}.
   We show that it is possible to impose this gauge in case of non-relativistic string as well. Then we are able to find Hamiltonian  on the reduced phase space. This Hamiltonian could be useful for the study of non-relativistic limit of some backgrounds as for example non-relativistic limit of $AdS_5\times S^5$. Similar problem was recently studied in 
 remarkable paper \cite{Fontanella:2021hcb}. It would be certainly nice to perform similar  analysis in case of new non-relativistic string action \cite{Bidussi:2021ujm}. We hope to return to this problem in future. 
 
 The structure of this paper is as follows. In the next section we review construction of new non-relativistic string action as was performed recently in 
 \cite{Bidussi:2021ujm}. Then in section (\ref{third}) we find Hamiltonian form of this theory. Finally in section (\ref{fourth}) we find gauge fixed form of this theory switching to uniform light-cone gauge.

\section{New Non-Relativistic String}\label{second}
In this section we review construction of recently proposed non-relativistic string
action  \cite{Bidussi:2021ujm}
\begin{equation}
S=-\frac{T}{2}\int d^2\sigma \sqrt{-\tau}[
\tau^{\alpha\beta}h_{MN}+\epsilon^{\alpha\beta}m_{MN}]\partial_\alpha x^M
\partial_\beta x^N \ . 
\end{equation}
Our goal is to show how such an action can be derived, following  \cite{Bidussi:2021ujm}. Let us introduce vielbein $e_M^{ \ \underline{a}}$ so that 
the target space metric has the form
\begin{equation}\label{gmunu}
g_{MN}=e_M^{ \ \ua}e_N^{ \ \ub}\eta_{\ua\ub}  \ , 
\end{equation}
where we use the similar notation as in  \cite{Bidussi:2021ujm} so that  frame indices are $\ua,\ub=0,\dots,D-1$ and where $\eta_{\ua\ub}=
\mathrm{diag}(-1,1,\dots,1)$. Note that space-time indices are $M,N=0,1,\dots,D-1$. Following 
\cite{Bidussi:2021ujm} we also introduce parametrization 
of NSNS two form $B_{MN}$ as
\begin{equation}
B_{MN}=\frac{1}{2}\eta_{\ua\ub}(e_M^{ \ \ua}\pi_N^{ \ \ub}-
e_N^{ \ \ua}\pi_M^{ \ \ub}) \ . 
\end{equation}
To begin with let us write
 Nambu-Goto form of the action for relativistic string in general background
\begin{equation}
S=-cT_F\int d^2\sigma \sqrt{-\det g_{\alpha\beta}}-cT_F
\int d^2\sigma \frac{1}{2}\epsilon^{\alpha\beta}B_{\alpha\beta} \ , 
\end{equation}
where $\epsilon^{01}=1=-\epsilon_{01}$ and $T_F$ is string tension.
Note that $M,N=0,1,\dots,D-1$. As in 
\cite{Bidussi:2021ujm} we introduce
indices $\ua=(A,a)$ corresponding to directions longitudinal and transverse to 
string world-sheet where $A=0,1$ are longitudinal and $a=2,\dots,D-1$ are transverse. 
Then we have
\begin{equation}
e_M^{ \ \ua}=(cE_M^{ \ A},e_M^{ \ a}) \ , \quad 
\pi_M^{ \ \ua}=(c\Pi_M^{ \ A},\pi_M^{ \ a}) \ 
\end{equation}
so that
\begin{eqnarray}\label{gBexp}
& &g_{\alpha\beta}=c^2 \eta_{AB}E_\alpha^{ \ A}E_\beta^{ \ B}+
\delta_{ab}e_\alpha^{ \ a}e_\beta^{ \ b}  \ ,  \nonumber \\
& &B_{\alpha\beta}=\frac{1}{2}c^2\eta_{AB}(E_\alpha^{ \ A}\Pi_\beta^{ \ B}-
E_\beta^{ \ A}\Pi_\alpha^{ \ B})+\frac{1}{2}\delta_{ab}(e_\alpha^{ \ a}
\pi_\beta^{ \ b}-e_\beta^{ \ a}\pi_\alpha^{ \ b}) \ . \nonumber \\
\end{eqnarray}
Then as in \cite{Bidussi:2021ujm}
we parametrize longitudinal components as
\begin{eqnarray}\label{EmuA}
E_M^{ \ A}=\tau_M^{ \ A}+\frac{1}{2c^2}\pi_M^{ \ B}\epsilon_B^{ \ A} \ , 
\quad
\Pi_M^{ \ A}=\epsilon^A_{ \ B}\tau_M^{ \ B}+\frac{1}{2c^2}\pi_M^{ \ A} \ , 
\nonumber \\
\end{eqnarray}
where $\epsilon_B^{\ A}=\epsilon_{BC}\eta^{CA}$. Inserting (\ref{EmuA}) into (\ref{gmunu})  we finally get
\begin{eqnarray}
& &g_{\alpha\beta}=c^2 \tau_{\alpha\beta}
+\frac{1}{2}\eta_{AB}(\tau_\alpha^{ \ A}\pi_\beta^{ \ C}\epsilon_{C}^{ \ B}+
\tau_\beta^{ \ B}\pi_\alpha^{ \ C}\epsilon_{C}^{ \ B})+h_{\alpha\beta}
\nonumber \\
& &+\frac{1}{4c^2}\eta_{AB}\pi_\alpha^{ \ C}\epsilon_C^{ \ A}\pi_\beta^{  \ D}
\epsilon_D^{ \ B} \ , \nonumber \\
\end{eqnarray}
where
\begin{equation}
\tau_{\alpha\beta}=\tau_\alpha^{\  A}
\tau_\beta^{ \ B}\eta_{AB} \ , \quad 
h_{\alpha\beta}=e_\alpha^{ \ a}e_\beta^{ \ b}\delta_{ab} \ . 
\end{equation}
Then it is easy to see that
\begin{eqnarray}
\sqrt{-\det g_{\alpha\beta}}=
%c^2
%\sqrt{-\det \tau_{\alpha\beta}}(1+\frac{1}{2c^2}\tau^{\alpha\beta}h_{\beta\alpha}+
%\frac{1}{2c^2}\tau^{\alpha\beta}\eta_{AB}\tau_\beta^{ \ A}\pi_\alpha^{ \ C}
%\epsilon_C^{ \ B})=
%\nonumber \\
%c^2\sqrt{-\det\tau_{\alpha\beta}}(1+\frac{1}{2c^2}\tau^{\alpha\beta}h_{\beta\alpha})+
%\frac{1}{2}\sqrt{-\det\tau_{\alpha\beta}}\tau^\alpha_{ \ B}\epsilon_C^{ \ B}\pi_\alpha^{ \ C}= \nonumber \\
c^2\sqrt{-\det\tau_{\alpha\beta}}(1+\frac{1}{2c^2}\tau^{\alpha\beta}h_{\beta\alpha})+
\frac{1}{2}\epsilon^{\alpha\beta}\tau_\beta^{ \ A}\eta_{AB}\pi_\alpha^{\ B} \ , 
\nonumber \\
\end{eqnarray}
where $\tau^{\alpha\beta}=\tau^\alpha_{ \ A}\tau^\beta_{ \ B}\eta^{AB}$ is $2\times 2$ matrix  inverse
to $\tau_{\alpha\beta}$. We also introduced $2\times 2$ twobein $\tau^\alpha_{ \ A}$ that obeys the condition 
\begin{equation}
\tau_\alpha^{ \ A}\tau^\beta_{ \ A}=\delta_\alpha^\beta \ , 
\quad
\tau_\alpha^{ \ A}\tau^\alpha_{ \ B}=\delta^A_B \ . 
\end{equation}
Explicitly we have
\begin{equation}
\tau^\alpha_{ \ A}=\frac{1}{\det \tau_\alpha^{ \ A}}
\left(\begin{array}{cc}
\tau_1^{ \ 1} & -\tau_0^{ \ 1} \\
-\tau_1^{ \ 0} & \tau_0^{ \ 0} \\ 
\end{array}\right)=-\frac{1}{\sqrt{-\det \tau_{\alpha\beta}}}
\epsilon^{\alpha\beta}\epsilon_{AB}\tau_\beta^{ \ B} \ , 
\end{equation}
where 
%$\epsilon^{01}=-\epsilon^{10}=1 \ , 
$\epsilon_{01}=-1=-\epsilon_{10}$.
Further, using (\ref{gBexp}) we obtain
\begin{eqnarray}
\frac{1}{2}\epsilon^{\alpha\beta}B_{\alpha\beta}=-c^2\sqrt{-\det \tau}+
\frac{1}{2}\epsilon^{\alpha\beta}\bB_{\alpha\beta} \ , 
\nonumber \\
\end{eqnarray}
where
\begin{equation}\label{bB}
\bB_{\alpha\beta}=\frac{1}{2}\delta_{ab}(e_\alpha^{ \ a}\pi_\beta^{ \ b}-
e_\beta^{ \ a}\pi_\alpha^{ \ b}) \ . 
\end{equation}
We further have
\begin{equation}
\tau^M_{ \ A}h_{MN}=0 \ , \quad 
\tau^M_{ \ A}\tau^N_{ \ B}\bB_{MN}=0 \ . 
\end{equation}
We see that two terms proportional to determinant of $\tau$ cancel each other. As a result we obtain non-relativistic action in the form 
\begin{equation}\label{Snonfinal}
S=-\frac{T}{2}\int d^2\sigma [\sqrt{-\tau}\tau^{\alpha\beta}h_{\alpha\beta}+
\epsilon^{\alpha\beta}m_{\alpha\beta}]  \ , 
\end{equation}
where we introduced rescaled tension  $cT_{F}=T$ and we have taken the limit $c\rightarrow \infty$. Finally we also defined $m_{
MN}$ as 
\begin{equation}
m_{MN}=\frac{1}{2}\eta_{AB}[\tau_M^{ \ A}\pi_N^{ \ B}-
\tau_N^{ \ A}\pi_M^{ \ B}]+
\frac{1}{2}\delta_{ab}[e_M^{ \ a}\pi_N^{ \ b}-e_N^{ \ a}\pi_M^{ \ b}] \ ,
\end{equation}
where $m_{\alpha\beta}=m_{MN}\partial_\alpha x^M\partial_\beta x^N$ is pull back of $m_{MN}$ to the world-sheet of the string. 
Before we proceed further it is instructive to stress following point. Using the fact that 
\begin{eqnarray}
-\frac{1}{2}\sqrt{-\det \tau}\tau^{\alpha\beta}
%\epsilon_{AB}(\tau_\alpha^{ \ A}\pi_\beta^{ \ B}+\tau_\beta^{ \ A}\pi_\alpha^{ \ B})=
%\nonumber \\
%=-\frac{1}{2}\epsilon_{AB}(\tau^\beta_{  \ C}\eta^{CA}\pi_\beta^{ \ B}+
%\tau^\alpha_{ \ C}\eta^{CA}\pi_\alpha^{ \ B})=\nonumber \\
%=\frac{1}{2}(\epsilon^{\beta\gamma}\tau_\gamma^{\ D}\eta_{DB}\pi_\beta^{ \ B}+
%\epsilon^{\gamma\alpha}\tau_\gamma^{ \ D}\eta_{DB}\pi_\alpha^{  \ B})=
%\nonumber \\
=\frac{1}{2}\epsilon^{\alpha\beta}(\tau_\alpha^{ \ A}\eta_{AB}\pi_\beta^{ \ B}-
\tau_\beta^{  \ A}\eta_{AB}\pi_\alpha^{ \ B})  \nonumber \\
\end{eqnarray}
it is easy to see that the action 
(\ref{Snonfinal})
can be rewritten into the form
\begin{equation}
S=-\frac{T}{2}\int d^2\sigma [\sqrt{-\tau}\tau^{\alpha\beta}\bh_{\alpha\beta}+
\epsilon^{\alpha\beta}\bB_{\alpha\beta}] \ , 
\end{equation}
where
\begin{equation}
\bh_{\alpha\beta}=\bh_{MN}\partial_\alpha x^M\partial_\beta x^N \ , \quad 
\bh_{MN}=h_{MN}-\frac{1}{2}\epsilon_{AB}(\tau_M^{ \ A}\pi_N^{ \ B}+
\tau_N^{ \ A}\pi_M^{ \ B}) \ . 
\end{equation}
If we identify $m_M^{ \ A}=-\frac{1}{2}\eta^{AB}\epsilon_{BC}\pi_M^{ \ C}$ we obtain standard form of non-relativistic action with an important difference that
$\bB$ is not arbitrary NSNS two form but it is explicitly defined in (\ref{bB}).

In this section we gave brief description  of procedure that was used in \cite{Bidussi:2021ujm} in order do find non-relativistic action (\ref{Snonfinal}). In the next section we proceed to the canonical analysis of this action.
\section{Hamiltonian Formalism}\label{third}
In order to find canonical form of the action  (\ref{Snonfinal}) it is useful to rewrite it into the form
\begin{equation}\label{Sactfirst}
S=-\frac{T}{2}\int d^2\sigma [\det\tau_\gamma^{ \ C}\tau^\alpha_{ \ A}\tau^\beta_{ \ B}\eta^{AB}h_{\alpha\beta}+
\epsilon^{\alpha\beta}m_{\alpha\beta}]  \ , 
\end{equation}
where we used the fact that $\sqrt{-\det \tau}=\det \tau_\gamma^{ \ A}$. From (\ref{Sactfirst})  we determine conjugate momenta
\begin{eqnarray}\label{defpM}
& &p_M=\frac{\partial \mL}{\partial(\partial_0 x^M)}=-\frac{T}{2}\tau_M^{ \ A}\tau^0_{ \ A}\det \tau_\gamma^{\ C}\tau^{\alpha\beta}h_{\alpha\beta}+T\det\tau_\gamma^{ \ C}
\tau^0_{ \ A}\eta^{AB}\tau^\beta_{ \  B}\tau_M^{ \ D}\tau^\alpha_{ \ D}h_{\alpha\beta}-
\nonumber \\
& &-T\det\tau_\alpha^{ \ D}\tau^{0\alpha}h_{\alpha M}-T m_{MN}\partial_1 x^N\nonumber \\
\end{eqnarray}
and where we used the fact that 
\begin{eqnarray}
& &\frac{\partial \det \tau_\gamma^{ \ C}}{\partial(\partial_0 x^M)}=
\frac{\partial (\tau_\alpha^{ \ A})}{\partial (\partial_0 x^M)}\tau^\alpha_{ \ A}
\det \tau_\gamma^{ \ C}=\tau_M^{ \ A}\tau^0_{ \ A}\det\tau_\gamma^{ \ C} \ , \nonumber \\
& & \frac{\partial \tau^{\alpha\beta}}{\partial (\partial_0 x^M)}h_{\alpha\beta}=
2\frac{\partial (\tau^\alpha_{ \ A})}{ \partial(\partial_0 x^M)}
\tau^\beta_{ \ B}\eta^{AB}h_{\alpha\beta}=-2\tau^0_{ \  A}\eta^{AB}
\tau^\beta_{ \ B}\tau_M^{ \ C}\tau^\alpha_{ \ C}h_{\alpha\beta} 
\nonumber \\
\end{eqnarray}
using
\begin{equation}
\frac{\partial (\tau^\alpha_{ \ A})}{\partial (\partial_0 x^M)}=
-\tau^\gamma_{ \ A}\frac{\partial \tau_\gamma^{ \ B}}{\partial (\partial_0 x^M)}
\tau^\alpha_{ \ B}=-\tau^0_{ \ A}
\tau_M^{ \ B}\tau^\alpha_{ \ B} \ . 
\end{equation}
Now with the help of the definition of the momenta $p_M$ given in (\ref{defpM}) 
we  determine bare Hamiltonian
\begin{eqnarray}
H_B=\int d\sigma (p_M\partial_0 x^M-\mL)=0
%\int d\sigma(-\frac{T}{2}
%\tau_0^{ \ A}\tau^0_{ \ A}\det \tau_\beta^{ \ B}\tau^{\alpha\beta}h_{\alpha\beta}+
%T\det \tau_\alpha^{ \ A}\tau^0_{ \ A}\eta^{AB}\tau^\beta_{ \ B}\tau_0^{ \ C}\tau^\alpha_{ \ C}h_{\alpha\beta}-\nonumber \\
%-T\det\tau_\alpha^{ \ A}\tau^{0\alpha}h_{\alpha 0}
%-Tm_{01}
%+
%\nonumber \\
%+\frac{T}{2}\det\tau_\alpha^{ \ A}\tau^\alpha_{ \ B}\tau^\beta_{ \ D}\eta^{AB}h_{\alpha\beta}+\frac{T}{2}
%\epsilon^{\alpha\beta}m_{\alpha\beta}=\nonumber \\
%\int d\sigma(-\frac{T}{2}
%\det \tau_\beta^{ \ B}\tau^{\alpha\beta}h_{\alpha\beta}+
%T\det \tau_\alpha^{ \ A}\tau^{0\alpha}h_{\alpha 0}-\nonumber \\
%-T\det\tau_\alpha^{ \ A}\tau^{0\alpha}h_{\alpha 0}
%+\frac{T}{2}\det\tau_\alpha^{ \ A}\tau^\alpha_{ \ B}\tau^\beta_{ \ D}\eta^{AB}h_{\alpha\beta}
%=0\nonumber \\
\end{eqnarray}
as expected. On the other hand from definition of momenta $p_M$ given in (\ref{defpM})  we obtain 
\begin{eqnarray}
\partial_1 x^Mp_M=
-\frac{T}{2}\tau_1^{ \ A}\tau^0_{ \ A}\det \tau_\gamma^{ \ D}
\tau^{\alpha\beta}h_{\beta\alpha}+T\det\tau_\gamma^{ \ D}\tau^{0\beta}\tau_1^{ \ C}
\tau^\alpha_{ \ C}h_{\alpha\beta}-T\det \tau_\gamma^{ \ D}\tau^{0\alpha}h_{\alpha 1}=0
\nonumber \\
\end{eqnarray}
and hence we have following constraint
\begin{equation}
\mH_\sigma=p_M\partial_\sigma x^M \approx 0 \ . 
\end{equation}
To proceed further we introduce $\Pi_M$ as
\begin{equation}
\Pi_M=p_M+Tm_{MN}\partial_1 x^N \ . 
\end{equation}
Then from (\ref{defpM}) we obtain
\begin{eqnarray}
& &\Pi_M \tau^M_{ \ A}\eta^{AB}\epsilon_{BD}\tau_1^{ \ D}=-\frac{T}{2}\tau^0_{ \ A}
\eta^{AB}\epsilon_{BC}\tau_1^{ \ C}\det \tau_\gamma^{ \ D}\tau^{\alpha\beta}h_{\beta\alpha}+
\nonumber \\
& &+T\det \tau_\gamma^{ \ D}\tau^{0\beta}
h_{\beta\alpha}\tau^\alpha_{ \ A}\eta^{AB}\epsilon_{BD}\tau_1^{ \ D}=
%=-\frac{T}{2}(\det \tau_\alpha^{ \ A})^2\tau^{00}\tau^{\alpha\beta}h_{\beta\alpha}
%+T(\det \tau_\alpha^{ \ A})^2\tau^{0\beta}h_{\beta\alpha}\tau^{\alpha 0}=\nonumber \\
%\frac{T}{2}(\det\tau_\alpha^{ \ A})^2\tau^{0\alpha}h_{\alpha\beta}\tau^{\beta 0}-
%\frac{T}{2}(\det\tau_\alpha^{ \ A})^2\det \tau^{\alpha\beta}h_{11}=
%\nonumber \\
\frac{T}{2}(\det\tau_\gamma^{ \ D})^2\tau^{0\alpha}h_{\alpha\beta}\tau^{\beta 0}
+\frac{T}{2}h_{11} \ .  \nonumber \\
\end{eqnarray}
Further, using (\ref{defpM}) we also find that 
 $\Pi_Mh^{MN}\Pi_N$ is equal to
\begin{equation}
\Pi_M h^{MN}\Pi_N=T^2(\det\tau_\gamma^{ \ D})^2\tau^{0\alpha}
\partial_\alpha x^M h_{MN}\partial_\beta x^N \tau^{\beta 0} \ . 
\end{equation}
Collecting these terms together we obtain following Hamiltonian constraint
\begin{eqnarray}\label{mHtau}
-2T\Pi_M \tau^M_{ \ A}\eta^{AB}\epsilon_{BD}\tau_1^{ \ D}+T^2h_{11}+\Pi_M h^{MN}\Pi_N\equiv \mH_\tau\approx 0 \ . \nonumber \\
\end{eqnarray}
In summary, we have theory with two primary constraints $\mH_\tau\approx 0  \ , 
\mH_\sigma\approx 0$. It is remarkable that the Hamiltonian constraint (\ref{mHtau}) has simpler form than in case of the Hamiltonian constraint that was 
found in case of non-relativistic string in stringy Newton-Cartan background
\cite{Kluson:2018grx}.

We conclude this section with the calculation of the Poisson brackets between constraints $
\mH_\sigma,\mH_\tau$ in order to show that the Poisson algebra is closed. It is convenient to introduce their smeared forms
\begin{equation}
\bT_\tau(N)=\int d\sigma N\mH_\tau \ , \quad 
\bT_\sigma(N^\sigma)=\int d\sigma N^\sigma \mH_\sigma \ . 
\end{equation}
First of all we have
\begin{eqnarray}
\pb{\bT_\sigma(N^\sigma),\bT_\sigma(M^\sigma)}=
\int d\sigma (N^\sigma \partial_\sigma M^\sigma-
\partial_\sigma N^\sigma M^\sigma)p_M\partial_\sigma x^M=
\bT_\sigma (N^\sigma \partial_\sigma M^\sigma-\partial_\sigma N^\sigma
M^\sigma)
\nonumber \\
\end{eqnarray}
that shows that this Poisson bracket vanishes on the constraint surface $\mH_\sigma\approx 0$. 
Further we have
\begin{equation}
\pb{\bT_\sigma(N^\sigma),\mH_\tau}=
-2\partial_\sigma N^\sigma \mH_\tau-N^\sigma\partial_\sigma \mH_\tau
\end{equation}
that again vanishes on the constraint surface $\mH_\tau\approx 0$. 
These results  show that $\mH_\sigma\approx 0$ is the first class constraint. Finally
we calculate Poisson bracket between smeared forms of the constraints $\mH_\tau$ 
\begin{eqnarray}
& &\pb{\bT_\tau(N),\bT_\tau(M)}=4\int d\sigma (\partial_\sigma MN-\partial_\sigma N M)
p_M(h^{MN}h_{NK}+\tau^M_{ \ A}\tau_K^{ \ A})\partial_\sigma x^K
\nonumber \\
& &=4\bT_\sigma(N\partial_\sigma M-M\partial_\sigma N) \nonumber \\
\end{eqnarray}
that again vanishes on the constraint surface $\mH_\sigma\approx 0$. 
In summary we find that the Poisson brackets of the primary constraints
vanish on the constraint surfaces $\mH_\sigma\approx 0 \ , \mH_\tau\approx 0$ and hence they are the first class constraints. 
\section{Uniform Light-Cone Gauge}\label{fourth}
In this section we study gauge fixed form of the theory whose Hamiltonian was
found in previous section. Our goal is  to impose uniform light-cone gauge. To do this we follow notation that was used recently in \cite{Frolov:2019nrr}. We firstly  select two coordinates  $x^0=t,x^{9}=\phi$ while 
$x^\mu\ , \mu,\nu=1,2,\dots,8$ label transverse directions. 
Two abelian isometries are realized by shift in $t$ and $\phi$.  As a consequence 
energy and momentum along these directions are conserved. We further presume that the
range of world-sheet coordinate $\sigma$ is $\sigma \in (-r,r)$ where $r$ will be
fixed by generalized light-cone gauge. Then the conserved energy and momentum along $\phi$ are given by following formulas
\begin{equation}
E=\int_{-r}^r d\sigma p_t \ , \quad J=\int_{-r}^r d\sigma p_\phi \ . 
\end{equation}
In order to impose uniform-light cone gauge let us introduce light-cone coordinates and momenta
\begin{eqnarray}
& &x^-=\phi-t \ , \quad x^+=\frac{1}{2}(\phi+t)+\alpha x^- \ , 
\nonumber \\
& &p_+=p_\phi+p_t \ , \quad p_-=\frac{1}{2}(p_\phi-p_t)-\alpha p_+ \ , \nonumber \\
\end{eqnarray}
with inverse relations
\begin{eqnarray}
& &\phi=x^++x^-(\frac{1}{2}-\alpha) \ , \quad t=x^+-x^-(\frac{1}{2}+\alpha) \nonumber \\
& & p_t=p_+(\frac{1}{2}-\alpha)-p_- \ , \quad p_\phi=p_-+p_+(\frac{1}{2}+\alpha) \ .
\nonumber \\
\end{eqnarray}
In principle we can insert these formulas into Hamiltonian constraint derived
in previous section however final expression is very complicated in the full
generality.
%\Let us insert these formulas into Hamiltonian constraint derived in previous section 
%and we get
%\begin{eqnarray}
%& &\mH_\tau=-2T\Pi_M \tau^M_{ \ A}\eta^{AB}\epsilon_{BD}\tau_1^{ \ D}+T^2h_{11}+\Pi_M h^{MN}\Pi_N=\nonumber \\
%& &=(p_t+m_{t\phi}\partial_\sigma\phi+Tm_{t\mu}\partial_\sigma x^\mu)
%\tau^t_{ \ A}\eta^{AB}\epsilon_{BD}(\tau_\mu^{ \ D}\partial_1 x^\mu+
%\tau_t^{ \ D}\partial_1 t+\tau_\phi^{ \ D}\partial_1\phi)-\nonumber \\
%& &-2T(p_\phi+m_{\phi t}\partial_1 t+m_{\phi\mu}\partial_1 x^\mu)
%\tau^\phi_{ \ A}\eta^{AB}\epsilon_{BD}(\tau_\mu^{ \ D}\partial_1 x^\mu+
%\tau_\phi^{ \ D}\partial_1 \phi+\tau_t^{ \ D}\partial_1 t)-\nonumber \\
%& &-2T(\tPi_\mu+Tm_{\mu\phi}\partial_\sigma\phi+
%T m_{\mu t}\partial_1 t)\tau^\mu_{ \ A}\eta^{AB}\epsilon_{BD}
%(\tau_\mu^{ \ D}\partial_1 x^\mu+
%\tau_\phi^{ \ D}\partial_1 \phi+\tau_t^{ \ D}\partial_1 t)+\nonumber \\
%& &+(\tPi_\mu+Tm_{\mu t}\partial_1t+T m_{\mu \phi}\partial_1\phi)
%h^{\mu\nu}(\tPi_\nu+Tm_{\nu t}\partial_1t+T m_{\nu \phi}\partial_1\phi)+\nonumber \\
%& &+2(\tPi_\mu+Tm_{\mu t}\partial_1t+T m_{\mu \phi}\partial_1\phi)h^{\mu t }+\nonumber \\
%+T^2 \partial_\sigma x^\mu h_{\mu\nu}\partial_\sigma x^\nu+
%T^2\partial_\sigma t h_{tt}\partial_\sigma t+
%T^2\partial_\sigma\phi h_{\phi\phi}\partial_\sigma\phi+
%+\nonumber \\
%& &+2T^2\partial_\sigma th_{t\mu}\partial_\sigma x^\mu+
%2T^2\partial_\sigma t h_{t\phi}\partial_\sigma\phi
%+2T^2\partial_\sigma\phi h_{\phi \mu}\partial_\sigma x^\mu \ , 
%\nonumber \\
%\end{eqnarray}
%We see that it is very difficult to analyse this situation in the full generality. 
For that reason we restrict to more tractable examples. Let us start with the
case of non-zero components of metric $h_{tt},h_{\phi\phi}$ where all off-diagonal components are zero. In other words the metric $h_{MN}$ has block diagonal form
\begin{equation}
h_{MN}=\left(\begin{array}{ccc}
h_{tt} & 0 & 0 \\
0 & h_{\phi\phi} & 0 \\
0 & 0 & h_{\mu\nu} \\ \end{array}\right) \ . 
\end{equation}
Clearly the first diagonal block has inverse $\mathrm{diag}(h_{tt}^{-1},h_{\phi\phi}^{-1})$ and consequently we find that 
$\tau_{t}^{  \ A}=\tau_\phi^{ \ A}=0$. Let us further presume that $m_{\mu\phi}=m_{\mu t}=0$ together with $m_{t\phi}=0$. Then the Hamiltonian constraint simplifies considerably
\begin{eqnarray}
& &\mH_\tau=
%-2T\tPi_\mu\tau^\mu_{ \ A}\eta^{AB}\epsilon_{BD}
%\tau_\mu^{ \ D}\partial_1 x^\mu
%+\tPi_\mu
%h^{\mu\nu}\tPi_\nu
%+p_t h^{tt}p_t+p_\phi h^{\phi\phi}p_\phi
%+\nonumber \\
%+T^2 \partial_\sigma x^\mu h_{\mu\nu}\partial_\sigma x^\nu+
%T^2\partial_\sigma t h_{tt}\partial_\sigma t+
%T^2\partial_\sigma\phi h_{\phi\phi}\partial_\sigma\phi =
\mH^{\bot}+p_+h^{++}p_++2p_+h^{+-}p_-+p_-h^{--}p_-+\nonumber \\
& &+T^2\partial_1 x^+h_{++}\partial_1 x^++2T^2\partial_1 x^+h_{+-}\partial_1 x^-+T^2\partial_1 x^-h_{--}\partial_1 x^- \ , 
\nonumber \\
\end{eqnarray}
where we defined
\begin{eqnarray}
 & &h^{++}=(\frac{1}{2}-\alpha)^2h^{tt}+(\frac{1}{2}+\alpha)^2h^{\phi\phi}, \quad   h^{--}=h^{tt}+h^{\phi\phi} \ , 
\nonumber \\
& &h^{+-}=(\frac{1}{2}+\alpha)h^{\phi\phi}-(\frac{1}{2}-\alpha)h^{tt} \ , 
\quad h_{--}=(\frac{1}{2}-\alpha)^2h_{\phi\phi}+(\frac{1}{2}+\alpha)^2h_{tt} \ , 
\nonumber \\
& &h_{++}=h_{tt}+h_{\phi\phi} \ , \quad 
h_{+-}=(\frac{1}{2}-\alpha)h_{\phi\phi}-(\frac{1}{2}+\alpha)h_{tt} \ , 
\nonumber \\
\end{eqnarray}
and where
\begin{eqnarray}
\mH^{\bot}=-2T\tPi_\mu\tau^\mu_{ \ A}\eta^{AB}\epsilon_{BD}
\tau_\mu^{ \ D}\partial_1 x^\mu
+\tPi_\mu
h^{\mu\nu}\tPi_\nu+T^2 \partial_\sigma x^\mu h_{\mu\nu}\partial_\sigma x^\nu \ , \quad \tPi_\mu=p_\mu+Tm_{\mu\nu}\partial_1 x^\nu \ . 
\nonumber \\
\end{eqnarray}
Now we are ready to impose uniform light-cone gauge by introducing following
gauge fixed functions \cite{Frolov:2019nrr}
\begin{equation}
\mG^+\equiv x^+-(\tau+a\frac{\pi}{r}m R_{\phi}\sigma) \ , \quad  \mG^-=p_--1 \ , \quad 
a=\frac{1}{2}+\alpha \ , 
\end{equation}
where $m$ is integer winding number that represents 
the number of times the string
winds around the circle parametrised by $\phi$. In fact, we implicitly presume that $\phi$ is angular variable $0\leq \phi\leq 2\pi R_\phi$ and hence
\begin{equation}
\phi(r)-\phi(-r)=2\pi m R_\phi \ . 
\end{equation}
Further, if we integrate $p_-=1$ over the whole string we can relate $r$ to the total momentum $P_-$ as 
\begin{equation}
r=\frac{P_-}{2} \ . 
\end{equation}
Clearly $\mG^+,\mG^-$ have non-zero Poisson brackets with $\mH_\tau,\mH_\sigma$ and hence form set of second class constraints that vanish strongly. As a result constraints $\mH_\tau=0,\mH_\sigma=0$  can be explicitly solved. We firstly solve $\mH_\sigma=0$ for $\partial_1 x^-$ and we get
\begin{eqnarray}
%\mH_\sigma=p_+\partial_\sigma x^++p_-\partial_\sigma x^-+p_\mu\partial_\sigma x^\mu=
%0
%\Rightarrow \nonumber \\
\partial_\sigma x^-=-p_\mu\partial_\sigma x^\mu-a\frac{\pi}{r}mR_\phi p_+ \ .
\nonumber \\
\end{eqnarray}
As we argued above  $\mH_\tau,\mH_\sigma$ vanish strongly and hence gauge fixed form of the action is equal to
\begin{eqnarray}
& &S=\int_{-r}^r d\tau d\sigma (p_
\mu\partial_\tau x^\mu+p_+\partial_\tau x^++p_-\partial_\tau x^-)=
\nonumber \\
& &=\int_{-r}^r d\tau d\sigma (p_\mu\partial_\tau x^\mu+p_+) \ . 
\end{eqnarray}
In other words we can identify Hamiltonian density on the reduced phase space as $
\mH_{red}=-p_+$ where $p_+$ is solution of the Hamiltonian constraint. Explicitly, it is solution of the equation
\begin{eqnarray}
& &0=\mH^{\bot}+p_+h^{++}p_++2p_+h^{+-}+h^{--}+\nonumber \\
& &+T_1^2\left(\frac{a\pi}{r}mR_\phi\right)^2h_{++}
-2T^2\left(\frac{a\pi}{r}mR_\phi\right)(p_\mu\partial_\sigma x^\mu+a\frac{\pi}{r}mR_\phi p_+)+\nonumber \\
& &T^2(p_\mu\partial_\sigma x^\mu-a\frac{\pi}{r}mR_\phi p_+)
h_{--}(p_\nu\partial_\sigma x^\nu+a\frac{\pi}{r}mR_\phi p_+) \ . 
\nonumber \\
\end{eqnarray}
This is quadratic equation for $p_+$ that can be  solved at least in principle. For simplicity let us consider the case of the string with zero winding so that $m=0$ and we obtain the result
\begin{eqnarray}
p_+=\frac{-h^{+-}-\sqrt{(h^{+-})^2-h^{++}(\mH^{\bot}+h^{--}+
		T^2(p_\mu\partial_\sigma x^\mu)^2h_{--})}}{h^{++}} \ . 
\end{eqnarray}
However the form of this expression suggests that this gauge is not well defied 
due to the presence of $-$ sign in front of $h^{++}$ under square root where $h^{++}$ is positive by definition.
On the other hand it is more natural to impose light cone gauge when we presume non-zero $\tau_t^{ \ A},\tau_\phi^{ \ A}$ together with $h_{tt}=h_{t\phi}=h_{\phi\phi}=0$. In other words $h_{MN}$ has non-zero components $h_{\mu\nu}$. We will also presume that the matrix $h_{\mu\nu}$ is non-singular and consequently we have $\tau_\mu^{ \ A}=0$. Then   the Hamiltonian constraint has the form 
\begin{eqnarray}
\mH_\tau=
%=-2T(p_t+m_{t\phi}\partial_\sigma\phi)
%\tau^t_{ \ A}\eta^{AB}\epsilon_{BD}(
%\tau_t^{ \ D}\partial_1 t+\tau_\phi^{ \ D}\partial_1\phi)-\nonumber \\
%-2T(p_\phi+m_{\phi t}\partial_1 t)
%\tau^\phi_{ \ A}\eta^{AB}\epsilon_{BD}(
%\tau_\phi^{ \ D}\partial_1 \phi+\tau_t^{ \ D}\partial_1 t)+\nonumber \\
%+\tPi_\mu
%h^{\mu\nu}\tPi_\nu+
%+T^2 \partial_\sigma x^\mu h_{\mu\nu}\partial_\sigma x^\nu=\nonumber \\
& &-2T(p_+(\frac{1}{2}-\alpha)-p_-+m_{t\phi}(\partial_\sigma x^++
\partial_\sigma x^-(\frac{1}{2}-\alpha)))
\tau^t_{ \ A}\eta^{AB}\epsilon_{BD}\times \nonumber \\
& &\times(\tau_t^{ \ D}(\partial_1 x^+-\partial_1 x^-(\frac{1}{2}+\alpha))
+\tau_\phi^{ \ D}(\partial_1 x^++\partial_1 x^-(\frac{1}{2}-\alpha))+\nonumber \\
& &-2T(p_-+p_+(\frac{1}{2}+\alpha)+Tm_{\phi t}
(\partial_1 x^+-\partial_1 x^-(\frac{1}{2}+\alpha))
\tau^\phi_{ \ A}\eta^{AB}\epsilon_{BD}\times \nonumber \\
& &\times(\tau_t^{ \ D}(\partial_1 x^+-\partial_1 x^-(\frac{1}{2}+\alpha))
+\tau_\phi^{ \ D}(\partial_1 x^++\partial_1 x^-(\frac{1}{2}-\alpha)))+\nonumber \\
& &+\tPi_\mu
h^{\mu\nu}\tPi_\nu+
+T^2 \partial_\sigma x^\mu h_{\mu\nu}\partial_\sigma x^\nu \nonumber \\
\end{eqnarray}
Now we are ready to fix the gauge. For simplicity we will presume string
with zero winding so that gauge fixing functions has the form
\begin{equation}
\mG^+\equiv x^+-\tau \ , \quad  \mG^-=p_--1 \ .
\end{equation}
Then from $\mH_\sigma=0$ we again get
 $\partial_\sigma x^-=-p_\mu\partial_\sigma x^\mu$
 and hence Hamiltonian constraint has the form 
\begin{eqnarray}
& &\mH_\tau=-2T(p_+(\frac{1}{2}-\alpha)-1+m_{t\phi}(1-p_\mu\partial_\sigma x^\mu
(\frac{1}{2}-\alpha)))\tau^t_{ \ A}\eta^{AB}\epsilon_{BD}\times \nonumber \\
& &\times 
(\tau_1^{ \ D}\partial_\sigma x^\mu p_\mu(\frac{1}{2}+\alpha)-
\tau_\phi^{  \ D}p_\mu\partial_\sigma x^\mu(\frac{1}{2}-\alpha))-
\nonumber \\
& &2T(1+p_+(\frac{1}{2}+\alpha)+
Tm_{\phi t}p_\mu\partial_\sigma x^\mu(\frac{1}{2}+\alpha))
\tau^\phi_{ \ A}\eta^{AB}\epsilon_{BD}\times \nonumber \\
& &\times 
(\tau_1^{ \ D}\partial_\sigma x^\mu p_\mu(\frac{1}{2}+\alpha)-
\tau_\phi^{  \ D}p_\mu\partial_\sigma x^\mu(\frac{1}{2}-\alpha))
+\mH^{\bot}=0
\nonumber \\
\end{eqnarray}
from which we can express $p_+$ as
\begin{eqnarray}
%2Tp_+((\frac{1}{2}-\alpha)\tau^t_{ \ A}+
%(\frac{1}{2}+\alpha)\tau^\phi_{ \ A})\eta^{AB}\epsilon_{BD}
%(\tau_1^{ \ D}\partial_\sigma x^\mu p_\mu(\frac{1}{2}+\alpha)-
%\tau_\phi^{  \ D}p_\mu\partial_\sigma x^\mu(\frac{1}{2}-\alpha))=
%\nonumber \\
%-2T[(-1+m_{t\phi}(1-p_\mu\partial_\sigma x^\mu
%(\frac{1}{2}-\alpha)))\tau^t_{ \ A}+
%(1+
%Tm_{\phi t}p_\mu\partial_\sigma x^\mu(\frac{1}{2}+\alpha))
%\tau^\phi_{ \ A}]
%\eta^{AB}\epsilon_{BD}\times \nonumber \\
%\times 
%(\tau_1^{ \ D}\partial_\sigma x^\mu p_\mu(\frac{1}{2}+\alpha)-
%\tau_\phi^{  \ D}p_\mu\partial_\sigma x^\mu(\frac{1}{2}-\alpha))
%+\mH^{\bot}\Rightarrow 
%\nonumber \\
& &p_+=-\frac{1}{(p_\mu\partial_\sigma x^\mu)
	((\frac{1}{2}-\alpha)\tau^t_{ \ A}+
	(\frac{1}{2}+\alpha)\tau^\phi_{ \ A})\eta^{AB}\epsilon_{BD}
	(\tau_t^{ \ D}(\frac{1}{2}+\alpha)-
	\tau_\phi^{  \ D}(\frac{1}{2}-\alpha))}\times \nonumber \\
& &\times([(-1+m_{t\phi}(1-p_\mu\partial_\sigma x^\mu
(\frac{1}{2}-\alpha)))\tau^t_{ \ A}+
(1+
Tm_{\phi t}p_\mu\partial_\sigma x^\mu(\frac{1}{2}+\alpha))
\tau^\phi_{ \ A}]
\eta^{AB}\epsilon_{BD}\times \nonumber \\
& &\times 
(\tau_1^{ \ D}\partial_\sigma x^\mu p_\mu(\frac{1}{2}+\alpha)-
\tau_\phi^{  \ D}p_\mu\partial_\sigma x^\mu(\frac{1}{2}-\alpha))
+\mH^{\bot}) \ . \nonumber \\
\end{eqnarray}
As we argued above $p_+=-\mH_{red}$ and hence we have derived
Hamiltonian density on the reduced phase space. We see that it has non-relativistic
nature since it is quadratic in momenta. This Hamiltonian density could
be starting point for further investigation of non-relativistic limit of some string theory backgrounds. 

{\bf Acknowledgment:}
\\
This work 
is supported by the grant “Integrable Deformations”
(GA20-04800S) from the Czech Science Foundation
(GACR).

\end{document}